\documentclass[proof]{WileyASNA-v1}

\articletype{Article Type}%

\received{17 March  2023}
\revised{}
\accepted{}

\begin{document}

\title{A brief History of Image Sensors in the Optical}

\author{Martin M. Roth}

\authormark{Roth}
%\authormark{Roth \textsc{et al}}

\address {\orgdiv{\\Leibniz-Institut f\"ur Astrophysik Potsdam (AIP)}, \orgaddress{\country{Germany}}}

\corres{Martin M. Roth\\ 
\email{mmroth@aip.de}}

\presentaddress{An der Sternwarte 16,\\ 14482 Potsdam,\\ Germany }

\abstract{Image sensors, most notably the Charge Coupled Device (CCD), have revolutionized observational astronomy as perhaps the most important innovation after photography. Since the 50th anniversary of the invention of the CCD has passed in 2019, it is time to review the development of detectors for the visible wavelength range, starting with the discovery of the photoelectric effect and first experiments to utilize it for the photometry of stars at Sternwarte Babelsberg in 1913, over the invention of the CCD, its development at the Jet Propulsion Laboratory, to the high performance CCD and CMOS imagers that are available off-the-shelf today.} 


\maketitle

% Example for bibliography citations:
% \citep{Beletic2003} see also
% \cite{Finger2003}.

\section{Introduction}\label{sec1}

\begin{figure}[t]	\centerline{\includegraphics[width=85mm]{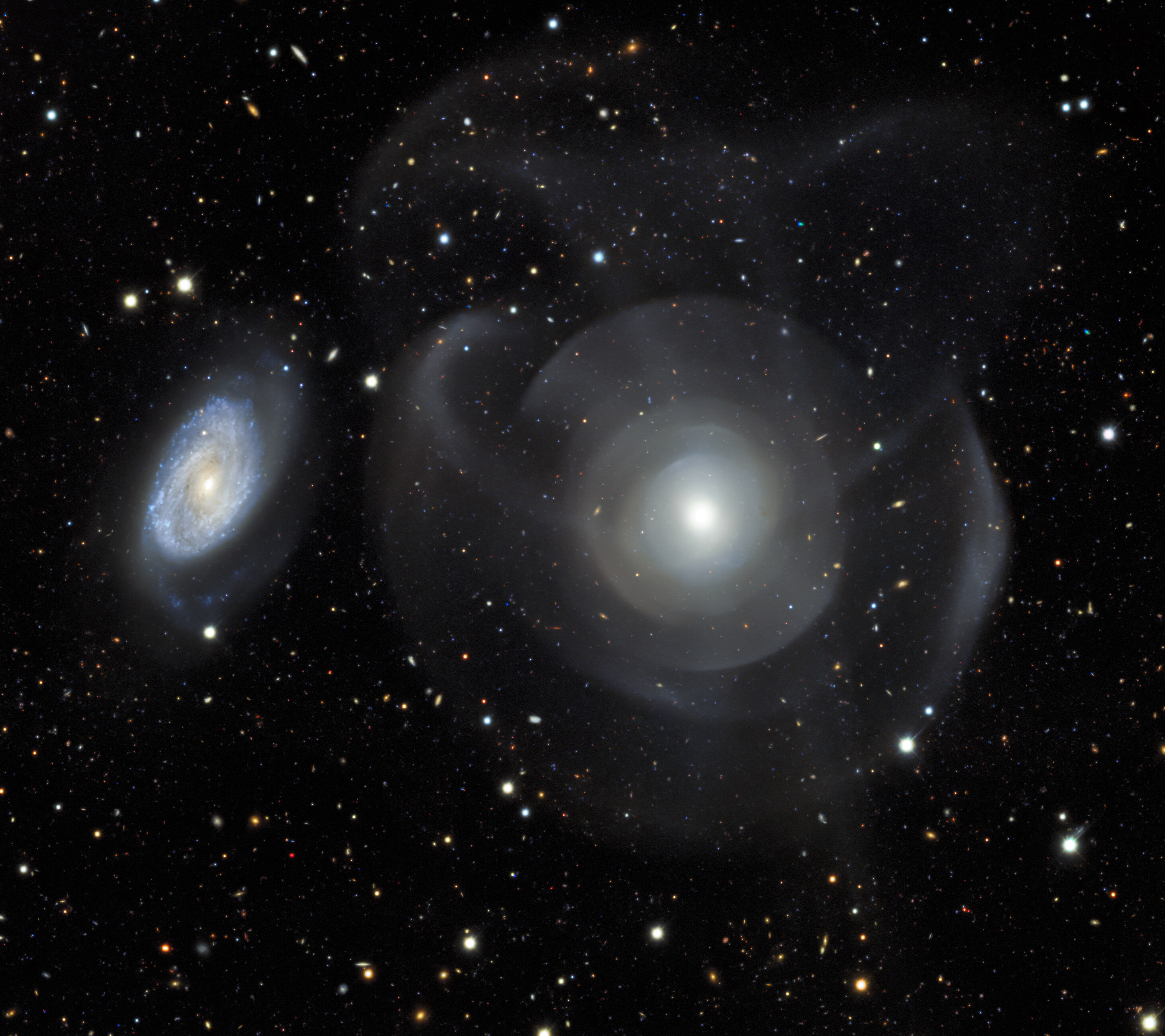}}
	\caption{CCD image of elliptical galaxy NGC\,474 with ring structure and companion spiral galaxy NGC\,470. Credit: DES/DOE/Fermilab/NCSA \& CTIO/NOIRLab/NSF/AURA.\label{fig1}}
\end{figure}

The Scientific Detector Workshop series (SDW), first held in 1996 in Garching
\citep{Beletic1998}, has spanned an era of exciting astronomical discoveries that would have been impossible without the ever growing light-collecting power of optical/near infrared telescopes on the ground and in space, and, equally importantly, the use of sensitive detectors, e.g. the vidicon, or the CCD. A few arbitrary examples for such discoveries may include the spectacular images from Neptune, taken by Voyager~2 in August 1989 \citep{Smith1989}, intricate details of the internal and faint outer structure of galaxies such as NGC\,474 \citep{Duc2015}, shown in Fig.~\ref{fig1}, or the famous Hubble Ultra Deep Field \citep{Beckwith2006}. The legacy of immensely productive programs like the Sloan Digital Sky Survey \citep{York2000} are unthinkable without CCD detectors, both for imaging, and spectroscopy. 

Younger generations of astronomers may take the existence of modern detectors for granted. However, long-term participants of SDW will recall the ubiquitous phrase ''{\it with the advent of CCDs in astronomy}'' in journal papers of the 1980's, that indicated a disruptive innovation whose impact on scientific progress in astronomy and astrophysics can hardly be over\-estimated. This {\it brief history of image sensors in the optical} is a sketchy attempt to highlight some of the important milestones of this success story. Hopefully it will be excused that in the interest of a focused and concise presentation, it cannot be complete and will inevitably neglect many important aspects of the technology, as well as of its application to science. 

The success story of near infrared detectors is presented in this volume by George Rieke (2023).

\section{The Photoelectric Effect }\label{sec2}

The end of the 19th century has seen a remarkable growth - if not an explosion - of research in natural sciences and technologies. To mention in passing, the venue of SDW2022 on the Telegrafenberg in Potsdam was chosen not the least to refer to this historical revolution in so far as the hill was once home to the Astrophysikalisches Observatorium Potsdam (AOP), founded in 1874. It was the first institute that picked up the term ''astrophysics'', to illustrate the then evolving transition from phenomenological observations and celestial mechanics in astronomy to the quantitative investigation of planets, stars, nebulae, and galaxies with the emerging understanding of atomic physics and refined instrumentation, first and foremost using spectroscopy as a new powerful tool - a mere 14 years after the fundamental paper about chemical analyses through spectroscopy by \cite{Kirchhoff1860} had been published.

Heinrich Hertz is perhaps more widely known for his pioneering work ''{\it Ueber sehr schnelle electrische Schwingungen}'' that led to the discovery of electromagnetic waves \citep{Hertz1887b}. In the same year, he published the paper ''{\it Ueber einen Einfluss des ultravioletten Lichtes auf die electrische Entladung}'' that is often quoted as the discovery of the photo\-electric effect \citep{Hertz1887a}. Shortly thereafter, \cite{Stoletow1888} published his work ''{\it Sur une sorte de courants electriques provoques par les rayons ultraviolets}'', followed by
\cite{Elster1889}, who reported their results  „{\it \"Uber die Entladung negativ electrischer K\"orper durch das Sonnen- und Tageslicht}". 

\begin{figure}[t]	\centerline{\includegraphics[width=85mm]{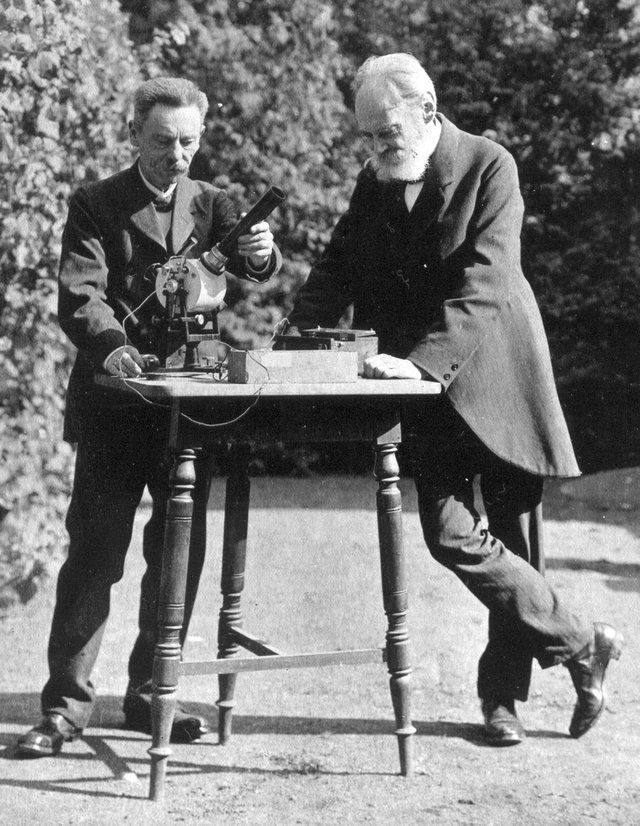}}
	\caption{Julius Elster and Hans Geitel experimenting in the garden of Elster's house. The instrument on the table is a photometer (credit: Archiv Fricke).\label{fig2}}
\end{figure}

Elster and Geitel (Fig.~\ref{fig2}) were highschool teachers, who refused to pursue an academic career, but as friends, living in the city of Wolfenb\"uttel, enjoyed to jointly perform experiments with electricity, radio activity, and other phenomena. It is interesting to note that as private researchers without an affiliation to an important university, they were the first ones to fabricate alkali photocells (patent obtainted in 1893), and based on these devices, to design and build an apparatus, whose purpose was to quantitatively measure the flux of a light source through the associated electrical current: the first photoelectric photometer was born, a single pixel light sensor \citep{Elster1889}. See also the detailed account of their biographies and achievements from \cite{Fricke2017}.

As a side note, it is also interesting to realize that somewhat later, Albert Einstein who was also concerned with the photoelectric effect, concluded that the quantization of light energy, associated with the {\it wavelength} of quanta, i.e. photons, suggests a statistical treatment for the description of the interaction of light with matter. This led to the statistical ''Einstein equations'' for the processes of absorption, emission, and stimulated emission that have become so ground-breaking for, e.g., the theory of radiative transfer in stellar photospheres, or the invention of the laser. It was the \cite{Einstein1905} paper for which he was awarded the Nobel Price for Physics in 1922.

In retrospect, it is perhaps fair to say that the two decades between \cite{Hertz1887a} and \cite{Einstein1905} have laid the foundation stones for what have become key enabling technologies in the second half of the 20th and in the 21st century. The enormous innovation potential for what would happen a hundred years later was likely not yet fully appreciated at the time.

\section{Introducing Photoelectric Photometry in Astronomy}\label{sec3}

Given the developments described above, one might expect that the use of Elster \& Geitel's photometer would have found immediate use as an objective measurement tool in astronomy. Indeed, 150 years ago, measuring the magnitude of a star was a highly subjective exercise of human observers, hence limited in accuracy. As a matter of fact, Elster \& Geitel themselves, although not being astronomers, tested their photometer on sky during a total solar eclipse in 1905 \citep{Elster1906}, and a total lunar eclipse in 1910 \citep{Elster1910}. At the same time, astronomers in the U.S.A. were experimenting with electrical photometers using selenium cells, that are based on a change of resistance as a function of illumination, rather than the photoelectric effect \citep{Stebbins1907,Stebbins1909,Stebbins1911}. However, as pointed out by \cite{Campbell1913}, a comparative study by \cite{Kemp1913} revealed that the photoelectric cell invented by Elster \& Geitel was ''{\it two hundred times more sensitive than the selenium cell}''.

\begin{figure}[t]	\centerline{\includegraphics[width=85mm]{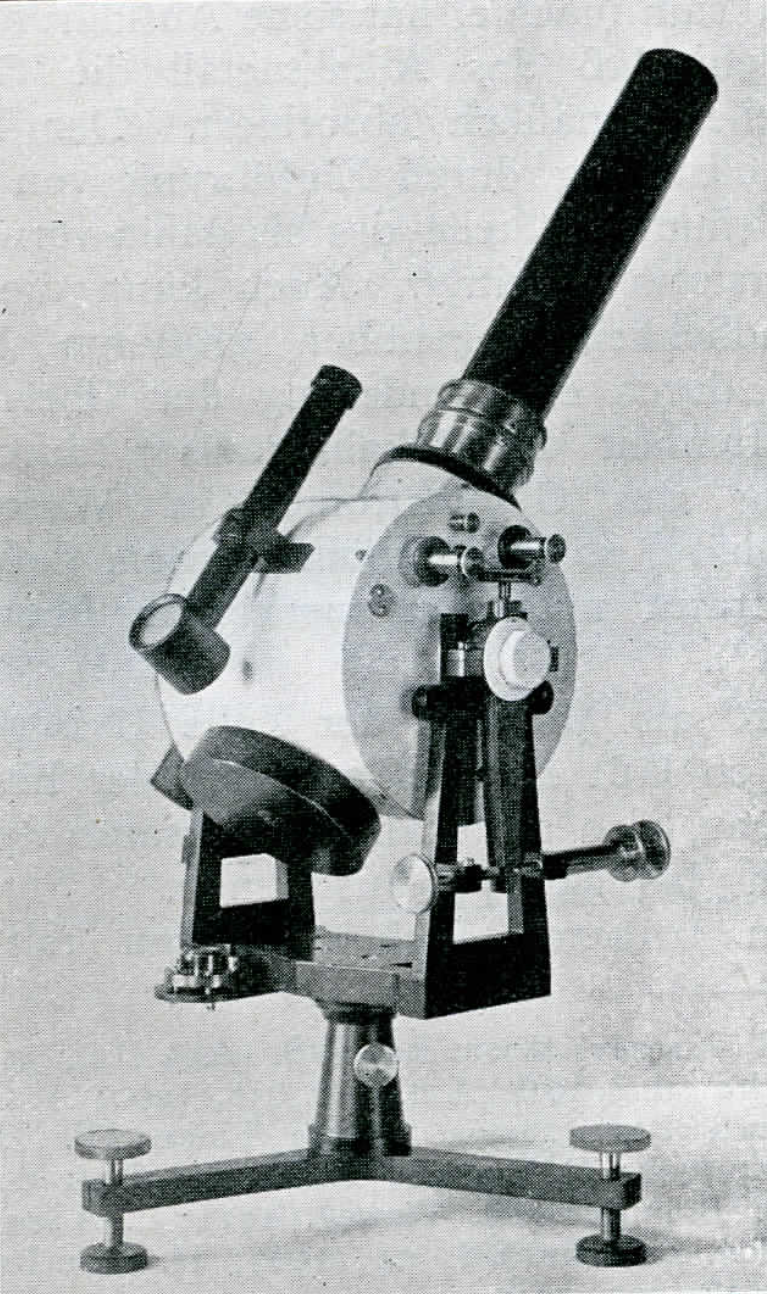}}
	\caption{Photometer developed by Elster \& Geitel (from \cite{Elster1912}).\label{fig3}}
\end{figure}

As observed by \cite{Walter1985}, the astronomical community at large was apparently as yet not prepared to adopt the novel technique. Paul Guthnick, astronomer at the Berlin Observatory, who was responsible for the removal to the new Babelsberg site near the city limits of Potsdam, that was finished in 1913, noted: ''{\it Last year I had the opportunity to carry out a plan which I had been nurturing for a number of years: using the photoelectric effect in alkali cells for astronomical measurements. It is well known that Prof. Elster and Prof. Geitel have tried for a long time to make astronomers realize the inestimable advantages which the use of such cells had to offer with regard to astronomical photometry. The apparent lack of interest which was expressed by the astronomical side can be probably explained only by the fact that the physical-technical difficulties to be overcome discouraged the non-physicists or made success itself appear questionable even at the outset.}'' The citation is a translation from \cite{Guthnick1913}, the first astronomical journal paper about photoelectric photometry with an instrument provided by Elster \& Geitel. Guthnick was obviously open-minded and knew about the achievements made by Elster and Geitel. He became director of the Babelsberg Observatory in 1921. The innovative power of the novel technique is probably best appreciated by citing again \cite{Campbell1913}, who wrote: ''{\it If the electric-cell photometer, attached to small telescopes, can measure the light of a fifth-magnitude star, in a few minutes, with an accuracy of a few thousandths of a magnitude, $\ldots$ we are entering a new era in the study of variable stars.}''

\begin{figure}[h!]	\centerline{\includegraphics[width=85mm]{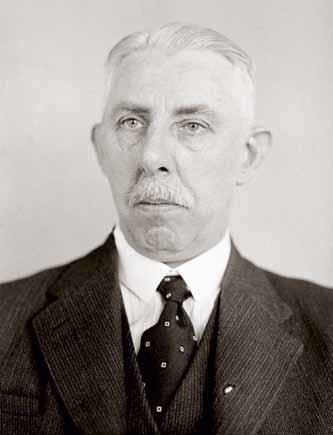}}
	\caption{Paul Guthnick, Director of Sternwarte Babelsberg and pioneer of photoelectric photometry in astronomy. © AIP.}\label{fig4}
\end{figure}

\begin{figure}[t!]	\centerline{\includegraphics[width=85mm]{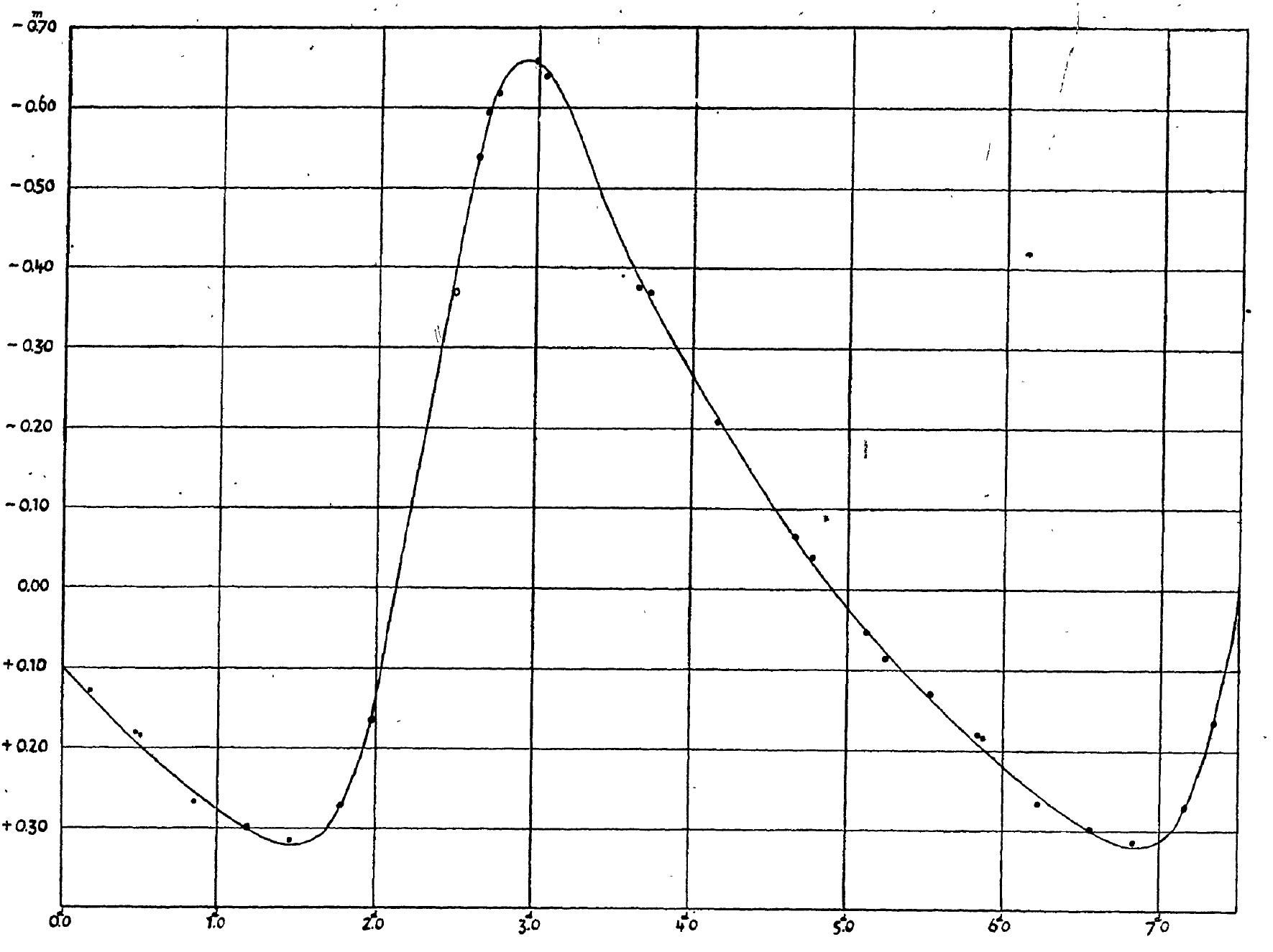}}
	\caption{Light curve of $\delta$ Cephei (from an article in Astronomische Nachrichten by \cite{Guthnick1919a}).}\label{fig5}
\end{figure}

Guthnick's own work on variable stars has been a good demonstration of this breakthrough. Fig.~\ref{fig5} shows the high quality of the fitted light curve to a total of 21 photometric observations of $\delta$ Cephei between October 1917 and January 1918. The data points deviate typically 0.005~mag from the fitted curve.

A following paper in the same volume \citep{Guthnick1919b} illustrates how his interpretation of the light curve, combined with the observation of in-phase radial velocity shifts, was happening at a time where the $\kappa$ opacity mechanism was not yet widely known and understood. However, it is remarkable that the era of photoelectric photometry began at the time of Henrietta Leavitt's discovery of the period-luminosity relation for Cepheids \citep{Leavitt1908}, hence the foundation for distance measurements that have changed our picture of the universe. A modern example of precision in the determination of the cosmic distance ladder is found in \cite{Riess2022}, with observations of Cepheid variables in the host galaxies of 42 Type Ia supernovae from more than 1000 orbits of the Hubble Space Telescope.

This section about photoelectric photometry would be incomplete without mentioning the dramatic sensitivity improvement of single-cell photometry by using secondary emission and the invention of the photomultiplier tube (PMT) \citep{Iams1935}. Originally driven by industry with the objective to develop television image tubes, the technology was quickly grasped by astronomers as soon as the first devices were becoming commercially available \citep{Kron1946}. 

The development of red sensitive PMTs and the definition of photometric pathbands like the UBV \citep{Johnson1953} and Kron-Cousins system \citep{Cousins1954,Kron1952} have revolutionized photometry and opened a new quantitative arena for astrophysics, e.g. the introduction of colors and the color-magnitude diagram, the study of variable stars, and many more. A full account of the topic would fill a book. PMTs are still in use in high energy astrophysics, such as the Pierre Auger Observatory, e.g. \cite{Aab2021}, IceCube \citep{Goldschmidt2001}, etc. However, in optical astronomy, they were soon outdated by the invention of the CCD.

\section{The Charge Coupled Device}\label{sec4}
The previous two sections were concerned with the {\it photoelectric} effect, and ''single pixel'' detectors without spatial resolution. However, there is also the {\it photovoltaic} effect, that was first reported by \cite{Becquerel1839}. The two are sometimes also called the {\it external} and {\it internal} photoelectric effects. While the former consists in the release of a free electron from a conductive layer of metal into vacuum  through the energy of an impinging photon, the latter is based on the elevation of an electron from the valence band to the conduction band of a semiconductor by absorbing the photon energy within the lattice of the material. Photovoltaic detectors are probably best known as photodiodes and phototransistors. 

As described in the book from \cite{Fisher2017}, the investigation of light-sensitive semiconductors was pursued at Bell Labs between the years of 1948 and 1952, and first reported by \cite{Shive1949,Shive1953}. However, for various reasons semiconductors were initially no serious competitors against PMTs for astronomical photometry. This situation changed completely when Williard S.\ Boyle and George E.\ Smith invented in 1969 the Charge Coupled Device, at Bell Labs as well. 

\begin{figure}[h]	\centerline{\includegraphics[width=90mm]{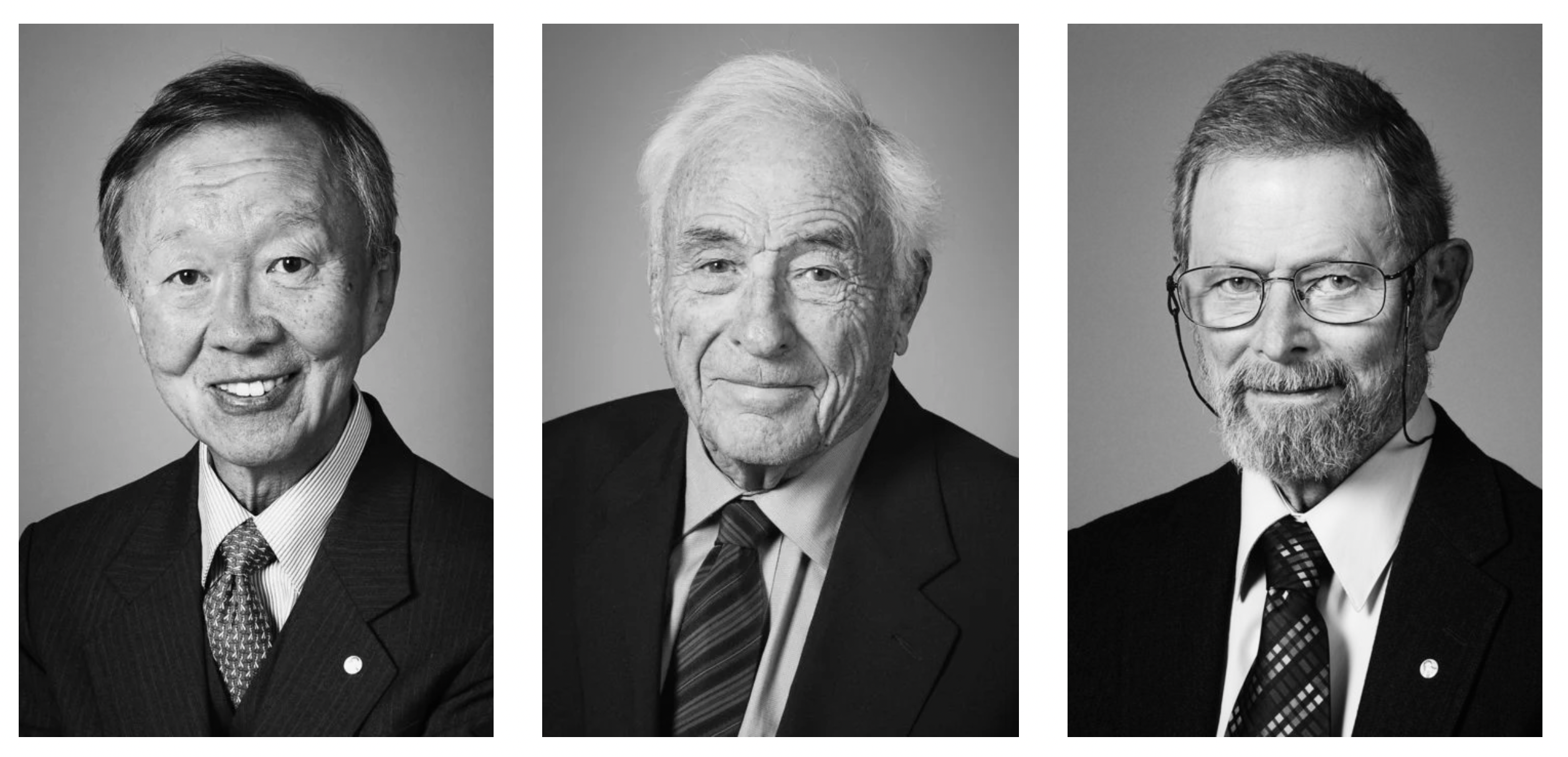}}
	\caption{Nobel Laureats for Physics in 2009 (from left to right): Charles K. Kao, Willard S. Boyle, George E. Smith. Kao was honored for his pioneering achievements with optical fibers, Boyle \& Smith for the invention of the CCD. 
 %Both have become key enabling technologies for astronomy.
    \\ © The Nobel Foundation, photo: U Montan.}\label{fig6}
\end{figure}

Together with Charles K.\ Kao,  who was honored for his achievements with the development of fiber optics, Boyle and Smith received the Nobel Price for Physics in 2009. They were given credit for essentially having launched a new era of digital imaging: ''{\it CCD image sensors have found important applications in many areas of society and science. They are found e.g. in digital cameras, scanners, medical devices, satellite
surveillance and in instrumentation for
astronomy and astrophysics. There are tens
of thousands of scientific publications that
mention the use of CCDs and many millions
of digital cameras that use CCD sensors.}'' \citep{Nobel2009}

Anyhow, the selection committee was probably not in the position to anticipate the even larger, ubiquitous diffusion of solid state imagers, in particular CMOS, in virtually all areas of society only a decade later, whether it would be used in smart phones, surveillance and web cameras, industrial process control, automotive, or many other applications. 

\begin{figure}[h]	\centerline{\includegraphics[width=70mm]{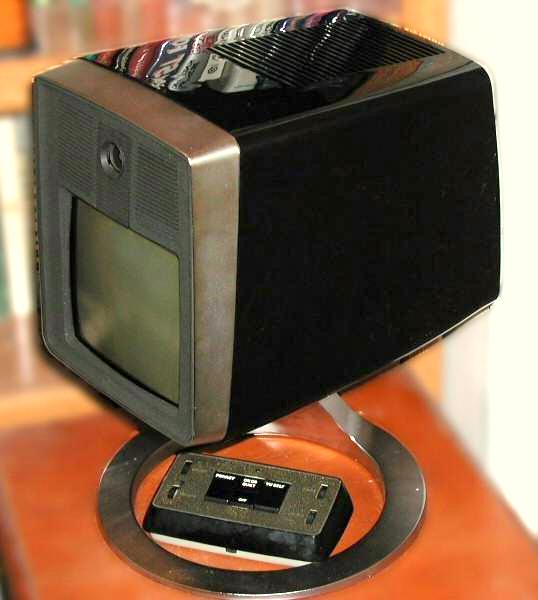}}
	\caption{AT\&T Bell Labs Picturephone Mod II from 1972 (Courtesy: LabguysWorld.com).}\label{fig7}
\end{figure}

However, the CCD imager was not a success story from the beginning. As \cite{Smith2010} pointed out in his Nobel lecture, an early driver for a compact semi-conductor based imager had been the ''Picturephone'', under development at AT\&T Bell Labs. The technology of the silicon diode array camera tube used for a first generation of devices to replace the vidicon tube, that had been ''notorious for its lack of robustness and longevity'', would be superseded by a solid state device, thus avoiding the inconvenient electron beam technology. An article by \cite{Hall1964} presented AT\&T's vision of the Picturephone with the expectation to become a huge market. Between 1964 and 1970, Bell Labs developed the Mod~I Picturephone, and an improved  version Mod~II thereafter (Fig.~\ref{fig7}. During a development phase of about 15 years, the company invested 500 million \$~US. However, on July 3rd, 1971, the New York Times reported in an article with the headline ''{\it Growth of Picturephones Disappoints Bell System}'' that customers had not accepted the new technology, and only a few systems were actually sold. Despite significant marketing efforts, even a prominent appearance in Stanley Kubrick's famous movie "2001: A Space Odyssey",
 the whole Picturephone project was eventually discontinued. In retrospect, it is an ironic turn of history that the commercial failure of an imaging telephone technology in the 1970's has turned today into a billion dollar market, thanks to the availability of smart phones and laptop computers with high quality CMOS cameras and high bandwidth internet communication. As a result, social media companies earn shareholder values that exceed the ones of some conventional industries.

By contrast, astronomy has engaged more successfully in the novel technology. Coming back to the Nobel Price awarded to Boyle and Smith in 2009, the accompanying material to the Nobel press release reads: ''{\it In 1974 the first image sensor had already been used to take photographs of the moon – the first astronomical images ever to be taken with a digital camera. With lightning speed, astronomers adopted this new technology; in 1979 a digital camera with a resolution of 320 x 512 pixels was mounted on one of the telescopes at Kitt Peak in Arizona, USA.}'' \citep{Nobel2009} 

Interestingly, the press release Scientific Background material quotes
under ''{\it Useful reading}'' the following references: \cite{Boyle1970}, \cite{Amelio1970}, and \cite{Janesick2001}. The reader is indeed referred to Jim Janesick's monograph that gives an excellent, comprehensive account of the historical development of the CCD technology, and the important role that the Jet Propulsion Laboratory (JPL) has played in it, but see also \cite{Janesick1992}. The JPL CCD development program was driven by the need for compact, robust, and reliable image sensors for scientific space missions. While the two Voyager missions were the last ones to fly vidicon detectors, the Solid State Imaging Experiment (SSI) onboard of Galileo, featuring
a front-side illuminated, virtual-phase, burried channel $800\times$800 pixels CCD from Texas Instruments \citep{Janesick2001}, was reported to exhibit a sensitivity a hundred times higher than the one of the NA and WA cameras of Voyager \citep{Belton1992}. From then on, the success story of CCDs unfolded vigorously. Perhaps the most prominent results known to the wider public are the spectacular images delivered by the Hubble Space Telescope (HST), since launch in 1990 for almost a quarter of century.

\begin{figure*}[th!]
\centerline{\includegraphics[width=175mm]
{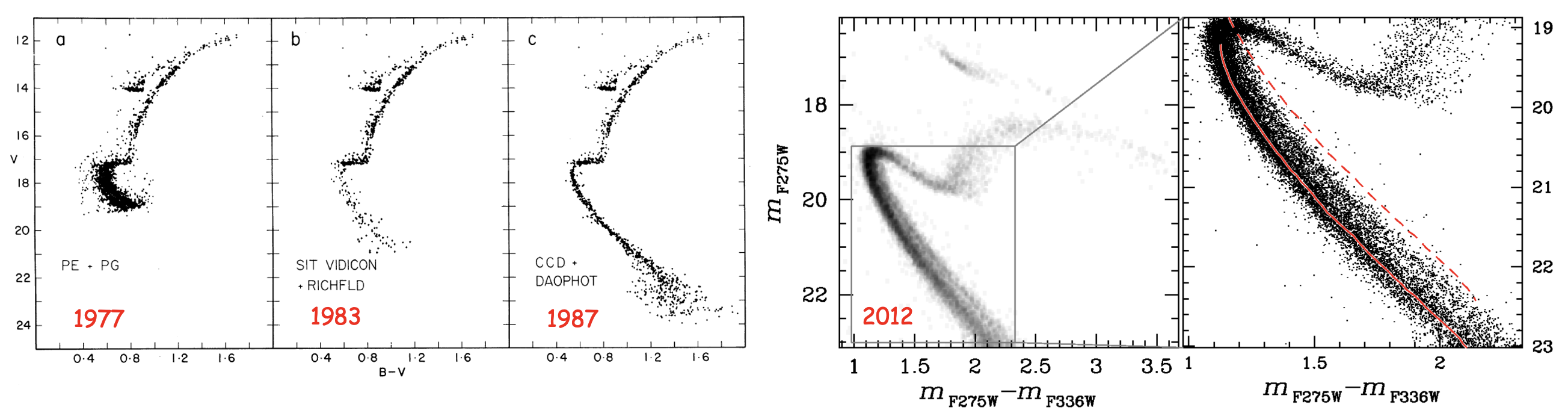}}\caption{Color magnitude diagram (CMD) for the globular cluster 47~Tuc. The three plots on the left, and two plots to the right, are reproduced from \cite{Hesser1987} and \cite{Milone2012}, respectively. Over a period of 35 years between 1977 and 2012 they illustrate the enormous progress in precision and accuracy that was enabled by detector technologies. 
\label{fig8}}
\end{figure*}

Insofar is it fair to conclude that while the time was not yet ripe for silicon image sensors to become a broad commercial success in the 1970's, the pressing need for scientific detector systems on the ground and in space has been the engine that has uniquely propelled the development of CCDs (and later scientific CMOS), to eventually become an innovation also on consumer markets as soon as they were established as efficient and reliable detectors for scientific applications. 

\section{Scientific Achievements with CCD Detectors}\label{sec5}

The progress made with CCD detectors for direct imaging and spectroscopy, also for polarimetry, could be illustrated with examples that would fill an entire book. Perhaps best known to a wider public are  spectacular HST images from planets, star clusters, dust clouds and gaseous nebulae, galaxies, and deep space exposures such as the Hubble Ultra Deep Field \citep{Beckwith2006}. 

\begin{figure}[bh!]	\centerline{\includegraphics[width=90mm]{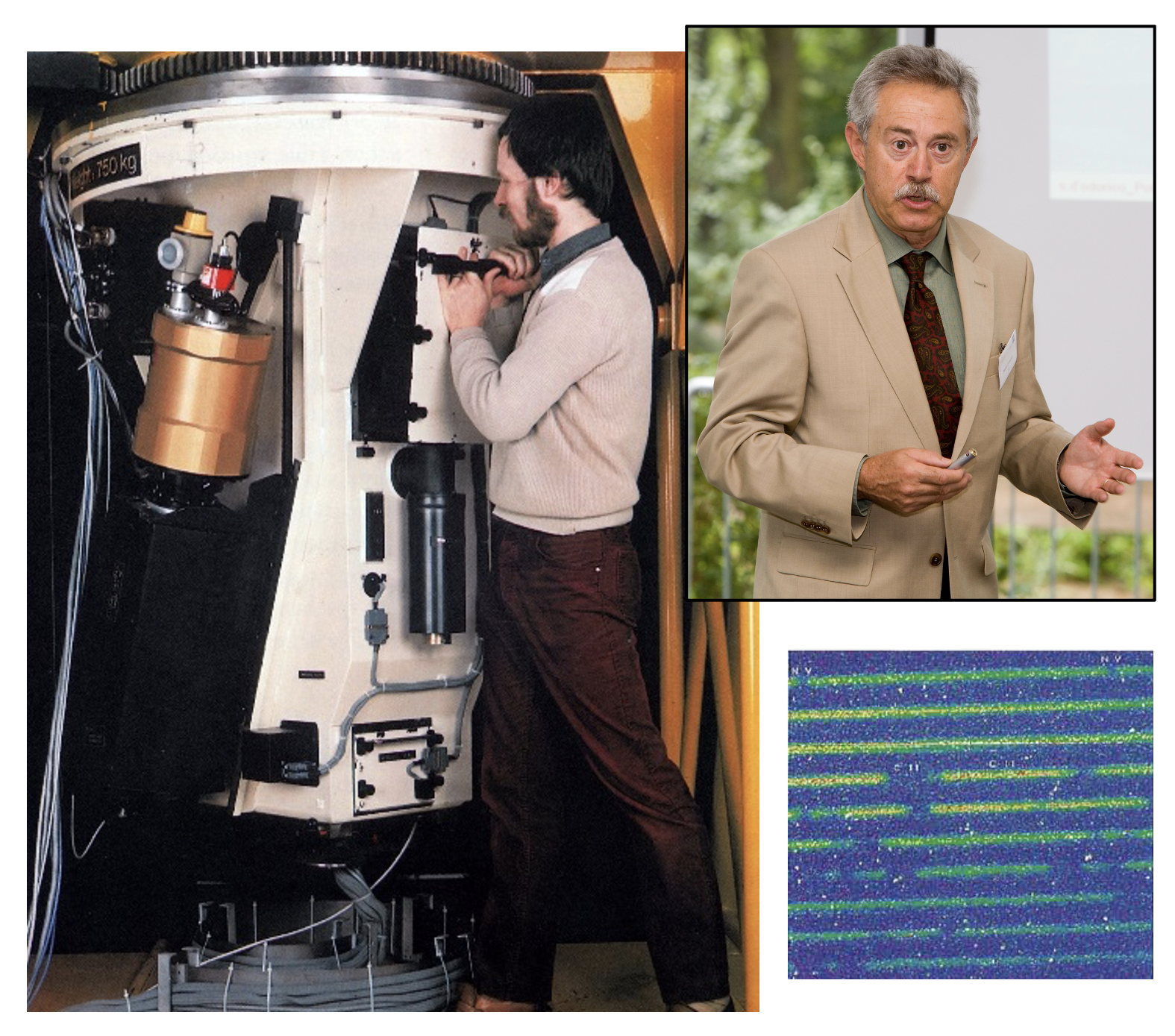}}
	\caption{The CCD program at ESO began with an RCA SID501 that was moved from direct imaging to the CASPEC spectrograph at the 3.6m Telescope \citep{DOdorico1983}, with a later upgrade to a larger chip. The insert shows a quasar absorption line spectrum \citep{Wampler1989}. The ESO CCD program was overseen by Sandro D'Odorico, who, in a sense, paved the way for the SDW series with a first workshop in 1986 \citep{ESO1987}. © ESO, AIP.}\label{fig9}
\end{figure}

Beyond aesthetic value, one may wish to pick an example for technical progress that has enabled a significant quantitative advance in science. Fig.~\ref{fig8} illustrates the improving precision and accuracy of the photometry of individual stars in the globular cluster 47~Tuc. Creating the CMD of a globular cluster from hundreds and thousands of cluster stars allows to compare temperature and luminosity of an entire stellar population with predictions from numerical models of stellar evolution, and to assess age and metallicity of the ensemble of stars. The sub-panels, labelled with years from 1977 to 2012 refer to photoelectric and photographic techniques (1977), over vidicon cameras (1983), to CCD photometry from the ground (1987), and finally high precision HST photometry from space (2012). It becomes immediately obvious that the sensitivity gain from photoelectric and photographic photometry to CCD photometry amounts to no less than 5 magnitudes (a factor of 100 in brightness), considering that the faintest measurable stars are found at 19th magnitude for the former, and 24th magnitude for the latter. Also the scatter on the main sequence in the x-axis direction (B-V color) has drastically reduced over time.

\begin{figure*}[th!]
\centerline{\includegraphics[width=175mm]
{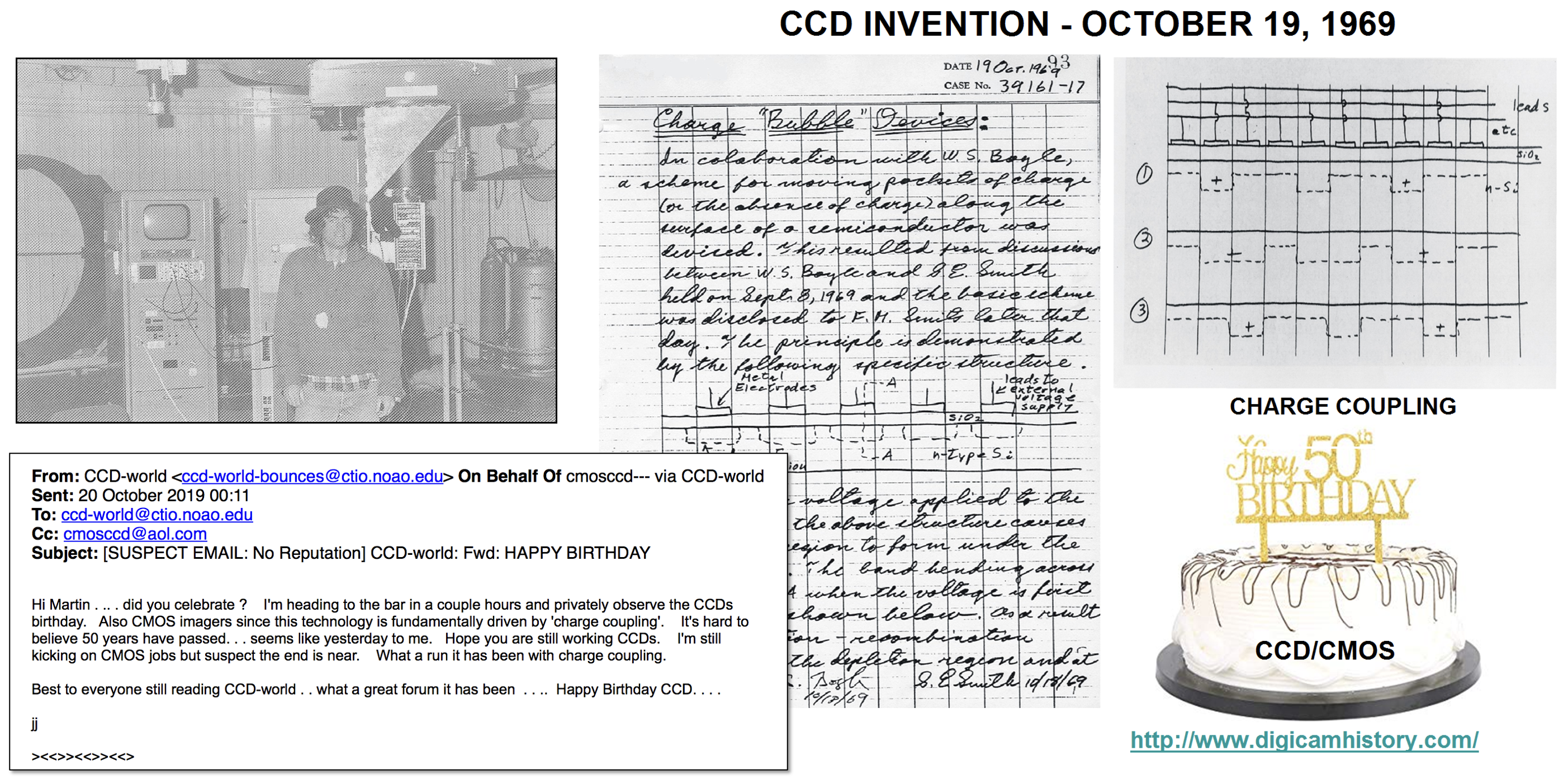}}\caption{The 50th anniversary of the CCD, celebrated via CCD-world. 
\label{fig10}}
\end{figure*}

This dramatic improvement has become possible for a number of reasons: first and foremost, CCDs offer a quantum efficiency (QE) that is two orders of magnitude higher than the one of photographic plates. Secondly, photography suffers from a non-linear response curve for the calibration from density to intensity, whereas most CCDs are linear all the way from the read noise floor to (almost) full well capacity, providing about three orders of magnitude larger dynamic range. Last not least, modern CCD cameras create immediately a digitally sampled output signal that can be directly stored and processed on a computer, while photographic plates have needed the extra step of scanning with a microdensitometer.

For the case presented in Fig.~\ref{fig8}, it worthwhile to mention that digital signal processing has been key to the advance of stellar population studies, such as in the globular cluster 47~Tuc that is presented here. The leftmost graph, with reference to the year 1977, illustrates that the techniques employed to perform photometry in crowded stellar fields on the basis of photoelectric aperture photometers (PE) and  photographic plates (PG) have suffered not only from poor QE and poor linearity, but also from crosstalk between the overlapping stellar images, that are heavily blended in the nuclear regions of any globular cluster. 

However, the linear recording of the point-spread-function (PSF) of bright template stars in less crowded regions of a CCD exposure has enabled the technique of PSF-fitting photometry that is capable of simultaneously deblending hundreds of stellar images over a large dynamic range, thus overcoming the crosstalk problem from previous techniques. The advance is clearly visible in Fig.~\ref{fig8} when comparing the two CMDs labelled 1977, and 1987.

This breakthrough has become possible in particular thanks to computer codes, e.g. the pioneering program DAOPHOT, that was developed by Peter Stetson at Herzberg Astrophysics, and that is still in use today \citep{Stetson1987}. At the time of writing, the paper has earned a total of 5006 citations according to the SAO/NASA Astrophysics Data System (ADS).

Further spectacular progress on globular cluster stars, and even resolved stellar populations in nearby galaxies \citep{Dalcanton2009} has become possible thanks to the superb angular resolution of HST. The example in Fig.~\ref{fig8}, right panel, illustrates that the accuracy of CCD photometry in 47~Tuc is so high that even the main sequence is breaking up into separate components, pointing to a complex star formation history and different stellar populations in the same cluster \citep{Milone2012}). Note that this CMD is composed on the basis of the HST filter set that is different from the Johnson B-V colors.

Obviously, the choice of this single example of CCD photometry is rather subjective, and one could make numerous other equally important, high scientific impact cases. Not only direct imaging, but also spectroscopy has experienced a boost of sensitivity and accuracy when CCDs were installed at spectrographs. Fig.~\ref{fig9} illustrates an example from ESO with the installation of an RCA chip of 320$\times$512 pixels in the CASPEC echelle spectrograph. In those days, exposure times of up to 5 hours were not uncommon, as the readout noise of these early devices was still high.

\section{The Scientific Detector Community}\label{sec6}

Rather than presenting technical detail, this brief historical review has focused on people. The author firmly believes that scientific progress is happening more than anything else because of the dedication and talent of enthusiastic individuals. The scientific detector community is a good example for this observation, however with the peculiarity that it is a mix of {\it academic} and {\it industrial} experts who have developed an amazing team spirit, despite the undeniable momentum of competition in the field. There is probably no better way to illustrate this through the moderated email list {\bf CCD-world}: a forum, that has enabled detector experts to share experience about technical challenges and trouble-shooting advice with their colleagues world-wide. The community is owing a lot to those who have spent their time to host this wonderful forum.
 
To quote Jim Janesick (priv. comm., 2023):
''{\it Yes.. .  it was the entire astronomical team that developed the CCD.   And indirectly CMOS since this technology is fundamentally the same as the CCD in terms of the solid state physics.   Over the course of many years . . .  I sent out daily handwritten memos to a couple hundred people who worked the CCD with the JPL.  This continued until CCD-World arrived along with e-mail.    A massive team that all wanted the same thing. . .    high performance and the ability to buy CCDs off the shelf.  The original memos are still stored in the basement of CIT.}'' 

Fig.~\ref{fig10} presents an example message that was sent via CCD-world on the occasion of the 50th anniversary of the invention of the CCD. The pictures, including a historical photograph of the sender, as well as a copy of the original lab book entry made at Bell Labs by Boyle and Smith on October 19th, 1969, are reproduced with permission from the book ''{\it Scientific charge-coupled devices}'' \citep{Janesick2001}. 

Beyond CCD-world, the communication and interaction within the detector community was further enhanced through the Scientific Detector Workshop series, that in essence began with the ESO-OHP workshop entitled ''{\it The optimization of the use of CCD detectors in astronomy}, held at Observatoire de Haute-Provence (St. Michel) in June 1986 \citep{ESO1987}, followed by the conference ''{\it CCDs in astronomy}'', held in Tucson in September 1989 \citep{ASPC1990}, the CCD Mini-workshop in Garching \citep{DOdorico1991}, and finally the SDW series in the format as it is known today (\cite{Beletic1998}, and following conferences). 

The compilation of workshops would remain incomplete without mentioning a somewhat elite conference for detector experts with {\it scuba diving skills}, organized by Bonner Denton on the Grand Cayman Island \citep{SPIE1992,Denton2000}.

\section{Summary and Conclusions}\label{sec7}

This brief history of optical image sensors is an attempt to feature some of the most remarkable milestones and highlights on the journey from the very first single pixel photodetector that was mounted to a telescope back in 1913, to the high performance large image sensors that have been assembled to mosaic focal planes of gigapixel formats of today. 

No attempt was made to be complete, which would have been beyond the scope of this article. Instead, the interested reader is, again, referred to the book ''{\it Scientific charge-coupled devices}'' \citep{Janesick2001} that not only presents a chapter about the history of the invention and development of the CCD, but also a comprehensive description of CCD properties and their characterization.

Also not attempt was made to cover topics such as CMOS sensors, the skipper, curved detectors, and other technologies that were presented at this workshop and elsewhere.

However, it is hoped that the thorny road of innovation, i.e. the process of converting an invention to a successful product on a market, has been illustrated well enough to convince the reader that high performance technologies for scientific applications, that do not necessary have the benefit of huge manufacturer profits, do not easily emerge per se. 

Instead, only a concerted effort of industry and academia with sufficient funding is capable of creating such products. In so far the huge upturn of knowledge in astronomy over the past five decades, that is sometimes called the ''{\it Golden Age of Astronomy}'' \citep{Pilachowsky2004}, has for a large part profited from the enormous progress with CCD imager development that was driven by science. The interesting experience of the Picturephone failure tells us that a technically brilliant idea is not necessarily sufficient to achieve a true innovation. Conversely, one can safely say that fundamental research, such as astronomy, can make every now and then an impact in terms of technology transfer, even though this may not have been the motivation for a technology development in the first place. As astronomers, we are lucky to benefit from such a constellation, thus the high performance sensors that are available off-the-shelf today.

\section*{Acknowledgments}
This work was supported by \fundingAgency{BMBF} under Contract No. \fundingNumber{03Z22AN11}. I am indebted to Regina von Berlepsch, Rudolf Fricke, Jim Janesick, and Sandro D'Odorico for sharing their historic resources. I also want to acknowledge the inspiration and energy of Jim Beletic and Paola Amico, who have been heart and soul of the Scientific Detector Workshop series over almost three decades.

\clearpage

\nocite{*}% Show all bib entries - both cited and uncited; comment this line to view only cited bib entries;
\bibliography{Wiley-ASNA}%

\end{document}